\documentstyle[psfig]{mn}
\newcommand{\be}{\begin{equation}}
\newcommand{\ee}{\end{equation}}
\newcommand{\ba}{\begin{eqnarray}}
\newcommand{\ea}{\end{eqnarray}}

\newcommand{\x}{\mbox{\boldmath $x$}}

\newcommand{\elb}{\mbox{\boldmath $\ell$}}
\newcommand{\thetab}{\mbox{\boldmath $\theta$}}
\newcommand{\lgl}{\langle}
\newcommand{\rgl}{\rangle}

\def\P3M{P$^3$M}
\def\and{, }
\font\mgn=cmti7
\def\eqnote#1{\marginpar{\mgn #1}}
\def\eqnote#1{{}}

\newfam\cyrfam
\font\tencyr=wncyr10 at 12pt
\font\sevencyr=wncyr7 at 8.5pt \font\fivecyr=wncyr5 at 6pt

\textfont\cyrfam=\tencyr \scriptfont\cyrfam=\sevencyr
            \scriptscriptfont\cyrfam=\fivecyr

\def\fracnum#1#2{\raise 2.1pt\hbox{$\scriptstyle #1$}\kern
-1.2pt/\kern -1.2pt \lower 2.1pt\hbox{$\scriptstyle#2$}\,}

\begin{document}


\title[Simulated power spectra] {Shear and magnification angular power
spectra and higher-order moments from weak gravitational lensing}

\author[Andrew J. Barber and A. N. Taylor] {Andrew J. Barber$^1$\thanks{Email:
A.J.Barber@sussex.ac.uk} and A. N. Taylor$^2$ \\
{}$^1$Astronomy Centre, University of Sussex, Falmer, Brighton, BN1
9QJ, U.K.\\
{}$^2$Institute for Astronomy, University of Edinburgh, Royal
Observatory, Blackford Hill, Edinburgh, U.K.
}

\date{Accepted 2001 ---. Received 2001 ---; in original form 2001 ---}

\maketitle

\begin{abstract}

We present new results on the gravitational lensing shear and
magnification power spectra obtained from numerical simulations of a
flat cosmology with a cosmological constant. These results are of
considerable interest since both the shear and the magnification are
observables. We find that the power spectrum in the convergence
behaves as expected, but the magnification develops a shot-noise
spectrum due to the effects of discrete, massive clusters and
symptomatic of moderate lensing beyond the weak-lensing regime. We
find that this behaviour can be suppressed by ``clipping" of the
largest projected clusters. Our results are compared with predictions
from a Halo Model-inspired functional fit for the non-linear evolution
of the matter field and show excellent agreement. We also study the
higher-order moments of the convergence field and find a new scaling
relationship with redshift. In particular, the statistic $S_3$ is
found to vary as $z_s^{-2.00\pm 0.08}$ (where $z_s$ is the source
redshift) for the cosmology studied, which makes corrections for
different median redshifts in different observational surveys
particularly simple to apply.

\end{abstract}


\section{INTRODUCTION}

Knowing the distribution and evolution of the large-scale structure in
the universe, together with the cosmological parameters which describe
it, are fundamental to obtaining a detailed understanding of the
cosmology in which we live. Studies of the effects of weak
gravitational lensing in the images of distant galaxies are extremely
useful in providing this information. In particular, since the
gravitational deflections of light arise from variations in the
gravitational potential along the light path, the deflections result
from the underlying distribution of mass, usually considered to be in
the form of dark matter. The lensing signal therefore contains
information about the clustering of mass along the line-of-sight,
rather than the clustering inferred from galaxy surveys which trace
the luminous matter.

Most obviously, weak lensing induces a correlated distortion of galaxy
images. The magnitude
of the correlations depends on the density parameter, $\Omega_m$, and
the value of the vacuum energy density parameter, $\Omega_V$, for the
universe, as these parameters reflect both the amount of mass and the
rate of evolution of structure. Consequently, the correlations depend
strongly on the redshifts of the lensed sources, as described by
Jain
\& Seljak (1997) and
Barber (2002). Recently a number of observational
results have been reported for the so-called cosmic shear signal,
which measures the variances in the shear on different angular
scales.
Bacon, Refregier \& Ellis (2000),
Kaiser, Wilson \& Luppino (2000),
Maoli
et al. (2001),
Van Waerbeke et al. (2000a,
b),
Wittman et al. (2000),
Mellier et al. (2001),
Rhodes, Refregier \& Groth (2001),
Van
Waerbeke et al. (2001),
Brown et al. (2002b),
Bacon et al. (2002),
Hoekstra, Yee \& Gladders (2002),
Hoekstra, Yee, Gladders, Barrientos, Hall \& Infante (2002) and 
Jarvis et al. (2002)
have all measured the cosmic shear and found
good agreement with theoretical predictions.

In addition to shearing, weak gravitational lensing may cause a source
at high redshift to become magnified or de-magnified as a result of
the amount and distribution of matter contained within the beam. The
degree of magnification is strongly related to the convergence, which
represents a projection of the density contrast, $\delta(x)$, at
position $x$ along the line of sight, and which is proportional to
$\Omega_m$. Consequently, measurements of statistics for the
convergence are able to provide cosmological constraints, and
comparisons of the power spectrum of the convergence with theoretical
predictions and numerical values serve to validate our theoretical and
numerical models  
(see, e.g., Moessner \& Jain, 1998, and 
Jain, 2002).

Of particular importance for interpreting weak lensing statistics is
the fact that the scales of interest lie largely in the
non-linear regime (see, e.g.,
Jain, Seljak \& White, 2000). On these
scales, the non-linear gravitational evolution introduces
non-Gaussianity to the convergence distribution, and this signature
becomes apparent in higher-order moments, such as the skewness. In
addition, the magnitude of the skewness values is very sensitive to
the cosmology, so that measurements of higher-order statistics in the
convergence may be used as discriminators of cosmology.

In this work, we have obtained weak lensing statistics from
cosmological $N$-body
simulations using an algorithm described by
Couchman, Barber \& Thomas
(1999) which computes the three-dimensional shear in the
simulations. The code has been applied to cosmological simulations
with $\Omega_m = 0.3$ and $\Omega_V = 0.7$; cosmologies of this type
will be referred to as LCDM cosmologies.  To obtain the required
statistics on different angular scales, the computed shear values have
been combined (using the appropriate angular diameter distance
factors and accounting for multiple deflections) along lines of sight
arranged radially from the observer's position at redshift $z =
0$. Detailed results are presented for background sources at 14
different redshifts ($z_s = 0.1$ to 3.6) and angular scales from
$1'$ to $32'$.

As a test of the accuracy of non-linear fits to the convergence power
we compare the numerically generated convergence power spectra with
our own theoretically predicted convergence spectra based on a Halo Model
fit to numerical simulations (Smith et al., 2002).
We also investigate the statistical properties of the magnification
power spectrum and test predictions of the weak lensing regime.
We also report on the expected redshift and scale dependence for
higher-order statistics in the convergence.

A brief outline of this paper is as follows.
In Section 2, we define the shear, reduced
shear, convergence and magnification in weak gravitational
lensing and outline how the magnification and convergence values are
obtained in practice from observational data.
In Section 3 we describe the relationships between the
power spectra for the convergence, shear and magnification
fluctuations, and how the power spectrum for the convergence relates
to the matter power spectrum. We also describe our methods for
computing the convergence power in the non-linear regime. Also in this
Section, the higher-order moments of the non-linear
convergence field are defined.
The numerical procedure we use to generate the
shear, convergence and magnification fields from the simulations
are presented in Section 4, while in Section 5 we present our results
for the numerical and
theoretical comparison of the convergence power spectra,
the power in the magnification fluctuations,
and the higher-order moments, particularly the $S_3$
statistic.
Finally we discuss the results and present our conclusions in
Section 6.

\section{Weak Lensing fields}

\subsection{Weak shear}

Ellipticity measurements of observed galaxy images can be
used to estimate the lensing shear signal.
One definition for the ellipticity (see, e.g.,
Blandford et al., 1991, and
Bartelmann \& Schneider, 2001)
is the complex ellipticity,
\begin{equation}
\epsilon = \frac{Q_{11}-Q_{22}+2iQ_{12}}{Q_{11}+Q_{22}+2(Q_{11}Q_{22}-
Q_{12}^2)^{\frac{1}{2}}},
\end{equation}
in which $Q_{ij}$ is the tensor of second brightness moments for a fixed
isophotal contour.

The ``reduced shear,'' $g$, for a galaxy image at angular position $\mbox{\boldmath$\theta$}$, is defined by
\begin{equation}
g(\mbox{\boldmath$\theta$}) \equiv
\frac{\gamma(\mbox{\boldmath$\theta$})}{1 -
\kappa(\mbox{\boldmath$\theta$})},
\label{g}
\end{equation}
where $\gamma$ is the complex shear and $\kappa$ is the lensing
convergence. Both $\gamma$ and $\kappa$ are obtained as projections
from the values of the second derivatives of the lensing potential
along the light path. Their detailed definitions and their relationships to
the lensing potential are given by 
Schneider, Ehlers \& Falco (1992)
and summarised by
Barber (2002).
The transformation between the source
ellipticities, $\epsilon^{(s)}$, and the image ellipticities,
$\epsilon$, is given by
\begin{equation}
\epsilon^{(s)} = \frac{\epsilon - g}{1-g^{\ast}\epsilon}
\label{source_ell}
\end{equation}
for $\mid g \mid \leq 1$. The asterisk in equation (\ref{source_ell})
denotes the complex conjugate. In the case of weak lensing, for which $\kappa$,
\mbox{$\mid \gamma \mid$} and \mbox{$\mid g \mid $} are much less than
unity, the
transformation reduces to
\be
    \epsilon \simeq \epsilon ^{(s)} + g
\ee
for low intrinsic source ellipticites.

The intrinsic ellipticities of given galaxies are not known, but
averaging over the binned galaxy distribution, and assuming random
ellipticities, yields a net lens shear:
\begin{equation}
\gamma \simeq g \simeq \langle \epsilon \rangle.
\end{equation}
This equality suggests that for weak lensing the variances in both the
shear and the reduced shear for a given angular scale are expected to
be similar. However, from numerical simulations,
Barber (2002)
has given explicit expressions for both as functions of redshift and
angular scale, which show the expected differences.

It is also possible to reconstruct the convergence from the shape
information alone, up to an arbitrary constant, using methods such as
those described by
Kaiser \& Squires (1993) and
Seitz \& Schneider (1996) for
the two-dimensional reconstruction of cluster masses.
Kaiser
(1995) generalised the method for applications beyond the linear regime.

Drawbacks to the reconstruction method arise from contamination by intrinsic
galaxy alignments 
(Pen, Lee \& Seljak, 2000, 
 Brown et al., 2002a,  
 Crittenden et al., 2001,
 Catelan, Kamionkowski \& Blandford, 2001,
 Mackey, White \& Kamionkowski, 2002, 
 Heavens, Refregier \& Heymans, 2000, and
 Croft \& Metzler, 2001), although these can be
statistically removed if redshift information is available
(
Heyman \& Heavens, 2002, and
 King \& Schneider, 2002). In addition,
map-making of the convergence field over a finite area suffers from
non-local effects  due to missing information beyond the survey
area
(Bacon \& Taylor, 2002). For these reasons it is useful to have an
alternative method for estimating the convergence.

\subsection{The magnification effect}

The lensing magnification, $\mu$, can be computed directly from
\begin{equation}
\mu =\left(|\det \cal A| \right)^{-1} =
\frac{1}{|(1-\kappa)^2-\gamma^2|},
\label{mu1}
\end{equation}
where $\cal{A}$ is
the two-dimensional Jacobian matrix which
describes the mapping of a source onto its image.
The effect of magnification on galaxy source counts results in a decrease
due to the increase in the lensed image area, and an increase in
counts due to the brightening of galaxies allowing their inclusion
in a flux-limited catalogue. For a power-law flux distribution,
the effect of lensing is 
(Broadhurst, Taylor \& Peacock, 1995, and 
 Bartelmann \& Schneider, 2001): 
\be
n_0(>S,z) \simeq \mu(z)^{\alpha-1} n(>S,z),
\ee
where $n(>S,z)$ and $n_0(>S,z)$ are the lensed and unlensed number of
galaxy images per unit solid angle with flux greater than $S$ and with
redshift within d$z$ of $z$, and $\alpha$ is the power-law exponent of $S$.
Hence with calibration of the underlying number count amplitude and slope
it is possible to estimate the magnification averaged over redshift.

Alternatively, magnification values may be obtained from the change in
image sizes at fixed surface brightness. This method is described
in detail by
Bartelmann and Narayan (1995) and summarised
concisely by
Bartelmann and Schneider (2001). See also Jain (2002) for a recent
discussion.

Estimates of the lensing magnification based on number counts suffer
from noise arising from the intrinsic clustering of the source
galaxies, if redshift information is not available (e.g., 
 Broadhurst, Taylor \& Peacock, 1995, and
 Bartelmann \& Schneider, 2001). With
redshift information one should either select galaxies at different
redshifts to remove the intrinsic clustering signal, or use the
brightening effect behind structure 
(Dye et al., 2001). Size
distortions may also be affected if size is correlated with
the environment. Again this may be removed by selecting galaxies at
different redshifts.

While the magnification signal-to-noise ratio is generally poorer than
that of the shear reconstruction method, it is valuable as an independent signal.
Furthermore, if good values for the convergence are available, then the
higher-order statistics are potentially very fruitful in
discriminating amongst cosmologies (see, e.g.,
Bernardeau, Van Waerbeke \& Mellier, 1997, and
Jain et al., 2000).

Finally, the convergence can be expressed (see
Jain et al., 2000, for example) as a projection of the density
contrast, $\delta(\mbox{\boldmath$\theta$},x_3)$:
\begin{equation}
2\kappa(\thetab,x_s) = \frac{3H_0^2\Omega_m}{2 c^2}\int_{0}^{x_s}
\! dx_3 \,  \frac{D(x_3)D(x_s-x_3)}{D(x_s)}
\frac{\delta(\mbox{\boldmath$\theta$},x_3)}{a(x_3)},
\label{projection}
\end{equation}
where $\thetab$ is the direction angle of the source at distance
$x_s$ along the coordinate direction $x_3$, $H_0$ is the Hubble
parameter, $D(x_3)$, $D(x_s-x_3)$ and $D(x_s)$ are the angular
diameter distances to position $x_3$, $x_s$ to $x_3$, and to
position $x_s$, respectively, and $a$ is the scale factor.

Although the lens convergence depends on the distance to the
source galaxy this dependence is usually lost by averaging over
the source distribution. However with redshift information the
full three-dimensional distribution of the density field can be fully
recovered 
(Taylor, 2001, and 
 Bacon \& Taylor, 2002).

\section{Statistical properties of lensing fields}

\subsection{The power spectra}

The convergence field can be expanded in two-dimensional Fourier modes on a flat-sky:
\be
    \kappa (\elb) = \int \! d^2 \theta  \, \kappa(\thetab) e^{i \elb . \thetab},
\ee
where $\elb$ is the angular wavenumber.
The two-point correlation of these modes defines the power spectrum,
$C^{\kappa \kappa}_\ell$:
\be
    \lgl \kappa(\elb) \kappa^*\!(\elb') \rgl = (2 \pi)^2
    C^{\kappa \kappa}_\ell     \delta_D(\elb -\elb').
\ee
Since the convergence can be expressed in the form of a
projection of the density contrast, the power spectrum
for the effective convergence can be obtained in
terms of the matter power spectrum, $P_{\delta}(k)$, for the density contrast
(e.g., Bartelmann \& Schneider, 2001):
\begin{equation}
C_\ell^{\kappa \kappa}=\frac{9H_0^4\Omega_m^2}{4c^4}\int_{0}^{w_H}\! dw \,
\left(\frac{\overline{W}(w)}{a(w)}\right)^2 P_{\delta}\left(\frac{\ell}{w},w\right),
\label{powerk}
\end{equation}
where the integral is evaluated over the comoving radial coordinate, $w$,
from 0 to the horizon, $w_H$, defined by
\be
    w_H = c \int \!\frac{dz}{H(z)},
\ee 
where $H(z)=H_0[(1+z)^3 \Omega_m + \Omega_V]^{1/2}$ and where we
have assumed a spatially flat universe; the weighting function,
$\overline{W}$, can be expressed as
\begin{equation}
\overline{W}(w) \equiv \int_{w}^{w_H} \!d w' \,
G(w')\frac{w'-w}{w'}, \label{pro}
\end{equation}
where $G(w)$d$w=p_z(z)$d$z$, where $p_z(z)$ is the matter
distribution function.

On angular scales smaller than about $10'$, the total power in the
effective convergence per logarithmic interval in wavenumber,
$\ell(\ell+1) C^{\kappa \kappa}_\ell/(2 \pi)$, is dominated by
galaxy clusters.
Jain et al. (2000) show that for the scales of interest in weak
lensing, $\ell(\ell+1)C^{\kappa\kappa}_\ell/(2 \pi)$ lies almost
entirely in this regime, and there is significant enhancement
(approximately an order of magnitude) of the power over linear
predictions on scales below $\ell \simeq 10^4$.

The two-point statistical properties
of the shear and convergence are closely related.
In the flat sky approximation the components $\gamma_1$ and
$\gamma_2$ of the shear are related to the effective convergence,
$\kappa$, in Fourier space (e.g., Barber, 2002):
\begin{equation}
\tilde{\gamma_1}^2(\elb) + \tilde{\gamma_2}^2(\elb) =
\tilde{\kappa}^2(\elb).
\end{equation}
Then it is clear that the power spectra for the shear,
$C^{\gamma\gamma}_\ell$, and the convergence,
$C^{\kappa\kappa}_\ell$, are the same in the case of weak lensing.

Finally, in the weak lensing regime, equation (\ref{mu1}) reduces
to \be
    \mu = 1 + \delta \mu \simeq 1 + 2\kappa,
\ee
 so that
\begin{equation}
C^{\delta \!\mu \delta \!\mu}_\ell \simeq  4C^{\kappa \kappa}_\ell
= 4C^{\gamma\gamma}_\ell, \label{powerm}
\end{equation}
where $C^{\delta \!\mu\delta \! \mu}$ is the power in the magnification
fluctuation, $\delta \mu$. This relation is questionable, since in general it is not obvious 
that the weak lensing regime will hold over the whole magnification
field. In particular, the denominator of the expression for $\mu$ in equation~(\ref{mu1}) 
is sensitive to the value of $\kappa$ near peaks in the convergence field.
We shall test this dependence in Section 5.2.

\subsection{Non-linear evolution of the matter field}

So long as the gravitational potential field is small the
perturbations in the matter density field can be fully non-linear.
The three-dimensional gravitational potential, $\Phi(\x)$, is related to
arbitrary perturbations in the mass-density field by Poisson's
equation,
 \be
 (\nabla^2 +3 H_0^2 \Omega_K )  \Phi(\x) = - \frac{3}{2} H_0^2 \Omega_m
    (1+z) \delta(\x,z),
 \ee
where $\Omega_K$ is the curvature energy-density parameter.

 A number of useful
techniques have emerged to allow us to transform from the linear
matter power spectrum to a fully non-linear spectrum. The first of
these originated from the work of
Hamilton et al. (1991) who
considered the evolution of the matter correlation function, and was
extended by
Peacock \& Dodds (1996) to account for the non-linear evolution of
the matter power spectrum. These methods are based on the
conservation of mass and a rescaling of physical lengths due to
gravitational collapse. These fits generally are accurate to
around the $10\%$ level.

More recently 
 Peacock \& Smith (2000) and
 Seljak (2000) have developed a model for the non-linear evolution of the power
spectrum based on the random distribution of dark matter haloes,
modulated by the large-scale matter distribution. This Halo Model
for non-linear evolution reproduces the matter power spectrum of
N-body simulations over a wide range of scales and has the
advantage of relating the linear and non-linear power at the same
scale. Given the utility of the non-linear fitting formula, 
 Smith et al. (2002) have presented a new set of fitting functions based
on the Halo Model functional form and calibrated to a set of $N$-body
simulations. These prove to be far more accurate ($\sim 1\%$)
than previous formul\ae. However these fits are only as accurate as
the underlying simulations used in the fitting, which in this case
were provided by the VIRGO
Consortium\footnote{http://star-www.dur.ac.uk/$\sim$frazerp/virgo/virgo.html}.
 In this paper we shall use the latter, more
accurate fits based on the Halo Model functional form. While 
Smith
et al. (2002) have compared their fit to the non-linear matter power
spectrum, here we compare for the first time the predicted
convergence power with the results of our simulations in Section 5.1.

\subsection{Higher-order moments in the convergence}

In addition to two-point statistics, higher-order statistical
properties of the lensing fields are also of interest as
non-linear evolution of the density field will introduce
non-Gaussianity (e.g., 
 Jain et al., 2000).
Bernardeau et al. (1997) have investigated
 analytically the dependence of higher-order moments in the
 convergence on the cosmological parameters, and in particular have
 discussed the ratio
\begin{equation}
S_3(\theta) \equiv \frac{\langle \kappa^3(\theta) \rangle}{\langle \kappa^2(\theta) \rangle^2}.
\label{s3}
\end{equation}
The significance of the $S_3$ statistic is that it is expected to be
independent of the normalisation of the power spectrum, and can also
be shown to be rather insensitive to the angular scale.

In the case of $\Omega_V=0$,
Bernardeau et al. (1997) have shown that
\begin{equation}
S_3(\Omega_m)\simeq -42\Omega_m^{-0.8}
\label{s3A}
\end{equation}
for $z_s \simeq 1$. The $\Omega_m$ dependence is slightly weaker for
sources at high redshift, and at low redshift $S_3$ becomes
approximately inversely proportional to $\Omega_m$. The redshift
dependence of $S_3$ in an Einstein-de Sitter cosmology is
approximately $z_s^{-1.35}$.

Jain et al. (2000) have investigated the values for $S_3$ in different
 cosmologies for sources at $z_s = 1$, including an LCDM cosmology using $N$-body simulations
 based on reconstructing the convergence values from the shear, and
 including the effects of noisy data. Using various statistics based
 on the reconstructed convergence, they show that there are clear
 differences in the $S_3$ values between the LCDM cosmology and an
 open cosmology, and also claim that $\Omega_m$ can be constrained to
 within an uncertainty of 0.1 -- 0.2 in a deep survey of several square degrees.
We shall study the $S_3$ statistic in more detail in Section 5.3.

\section{NUMERICAL PROCEDURE}

To evaluate the weak lensing statistics, we have applied the algorithm
for computing the shear in three dimensions, as described by
Couchman et al. (1999) to the cosmological $N$-body simulations
of the Hydra
Consortium\footnote{(http://hydra.mcmaster.ca/hydra/index.html)}
produced using the `Hydra' $N$-body hydrodynamics code
(Couchman, Thomas \& Pearce, 1995).  Simulations of the LCDM Dark
Matter only cosmology were used with $\Omega_m = 0.3,$ $\Omega_V =
0.7,$ power spectrum shape parameter $\Gamma = 0.25$ and
normalisation, $\sigma_8$, on scales of $8h^{-1}$Mpc of 1.22. The number of particles, each
of mass $1.29 \times 10^{11}h^{-1}$ solar masses, was $86^3$ and the
minimum value of the (variable) particle softening was chosen to be
$0.0007(1+z)$ in box units. The simulation volumes had comoving side
dimensions of 100$h^{-1}$Mpc.  To avoid obvious structure correlations
between adjacent boxes, each was arbitrarily translated, rotated (by
multiples of $90^{\circ}$) and reflected about each coordinate axis,
and in addition, each complete run was performed 10 times.

The general procedure for establishing the locations within the simulations
for the computations of the shear and for computing the values of the
elements of the shear matrices, is as described by
Barber (2002), with the multiple lens plane theory being applied along
the lines of sight. In this work, a total of $455 \times 455$ lines of sight were
used and 300 evaluation locations for the three-dimensional shear along each line of sight in each
simulation volume, thereby allowing regular sampling of the $2.6^{\circ} \times 2.6^{\circ}$ field of
view. With this number of lines of sight, the angular resolution
equates to the minimum value of the particle softening at the optimum
redshift, $z = 0.36$, for lensing of sources at a redshift of 1. To allow
for the larger angular size of the minimum softening at low redshifts
and also for the range of particle softening scales above the minimum
value, a resolution limit of 1 arcminute has been adopted for the data
analyses.

A total of 14 source redshift slices were selected to give good
statistical coverage of the redshifts of interest. These were
redshifts of $z_s = 0.10,$ 0.21, 0.29, 0.41, 0.49, 0.58, 0.72,
0.82, 0.88, 0.99, 1.53, 1.97, 3.07 and 3.57,
corresponding to the redshifts of the simulation boxes. Hereafter, we
shall only quote these redshifts to one decimal place for brevity.

From the statistics computed in each of the 10 simulation runs, we
computed the variances, the skewnesses, the statistic $S_3$, and the
power spectra for each of the source redshifts. The two-point and
higher-order moments were computed on angular scales of $1'.0$,
$2'.0$, $4'.0$, $8'.0$, $16'.0$ and $32'.0$ using a top-hat filter, and the
power spectra values were computed for a set of 15 wavenumber bins,
spaced logarithmically. Here we report on the results for these
statistics as computed for the effective convergence and the
magnification fluctuation (i.e., the departure of the magnification
value from unity). We computed these statistics from the convergence
values directly, rather than on convergence values reconstructed
from the shear.

For the power spectra, the square of the absolute values of the
Fourier transform of the convergence or magnification fluctuation were
normalised by multiplication by $\frac{L^2}{(2\pi)^2}.2\pi
\ell^2$, where $L$ is the
angular length of the map side in radians. With this definition of
the power spectrum, the results are essentially equivalent to
$\ell(\ell+1)C^{\kappa \kappa}_\ell/(2 \pi)$.

The computed values for the required statistics from each of the
$N=10$ runs were averaged, and the errors on the means of
$1\sigma/\sqrt{N}$ determined. However, since we found the power
in the magnification fluctuation to be a very noisy statistic, we
have determined the median values and computed the errors on the
medians for $C^{\delta \!\mu\delta \! \mu}_\ell/4$.

\section{RESULTS}

\subsection{Comparison of the convergence power spectra with
theoretical predictions.}

We have verified from our simulation results the equivalence of the
shear and convergence power spectra. The shear statistics, in
particular the shear variance, and
their redshift dependence are described fully in
Barber
(2002). Here we present our results for the convergence power spectra
for sources at the different redshifts obtained from our weak lensing
simulations and from the linear and non-linear theory predictions.

\begin{figure}
$$\vbox{
\psfig{figure=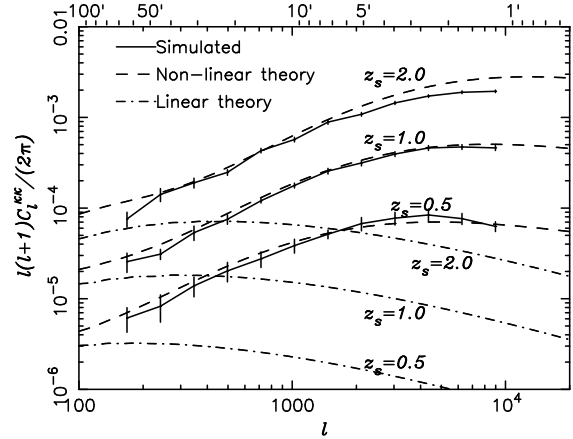,width=8.7truecm,angle=270} }$$
\caption{ Simulated $\ell(\ell+1)C^{\kappa\kappa}_\ell/(2\pi)$ values
(full lines) together with the linear (dot-dashed lines) and
non-linear (dashed lines) predictions  based on the Smith et al.
(2002) Halo Model-inspired fitting formula, for $z_s = 0.5$
(lowest set of curves), 1.0 (middle set) and 2.0 (uppermost set).
} \label{andyerror}
\end{figure}

The convergence power spectra values from the weak lensing simulations
are available for all the selected source redshifts.  Figure
\ref{andyerror} shows a comparison of the power spectra from our
simulations with linear theory and the non-linear convergence power
based on the Halo Model-inspired fitting functions of Smith et
al. (2002) for sources at redshifts 0.5, 1.0 and 2.0. We see that our
numerical results lie primarily in the non-linear regime and that
there is good agreement between the non-linear predictions and our
simulations. Notable discrepancies are apparent only for the highest
redshifts and in the highly non-linear regime, where we expect the
variance in the convergence field to be greatest. In terms of the
simulation data for redshifts greater than 1, our field of view is
effectively smaller because we make use of the periodicity in the
simulations when lines of sight extend outside the
volumes. Consequently, we sample in real terms a more limited area of
sky, which suggests that the quoted error bars for the high redshift
data may be understated. For the Halo-model, we simply note that its
predictions may not adequately account for the presence of voids and
detailed structure in the dark matter halos which become increasingly
important in terms of their gravitational lensing effects for high
redshift sources.








\subsection{The power in the magnification fluctuations}

\begin{figure}
$$\vbox{
\psfig{figure=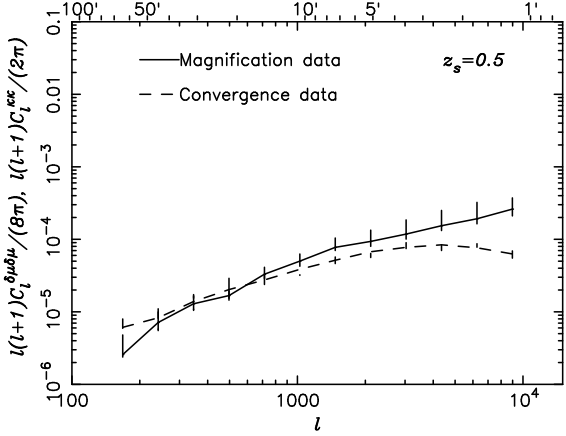,width=8.7truecm,angle=270} }$$ \caption{
Median $\ell(\ell+1)C^{\delta\!\mu\delta\! \mu}_\ell/(8 \pi)$ and
mean $\ell(\ell+1)C^{\kappa\kappa}_\ell/(2 \pi)$ for $z_s = 0.5$. }
\label{pm05}
\end{figure}

\begin{figure}
$$\vbox{
\psfig{figure=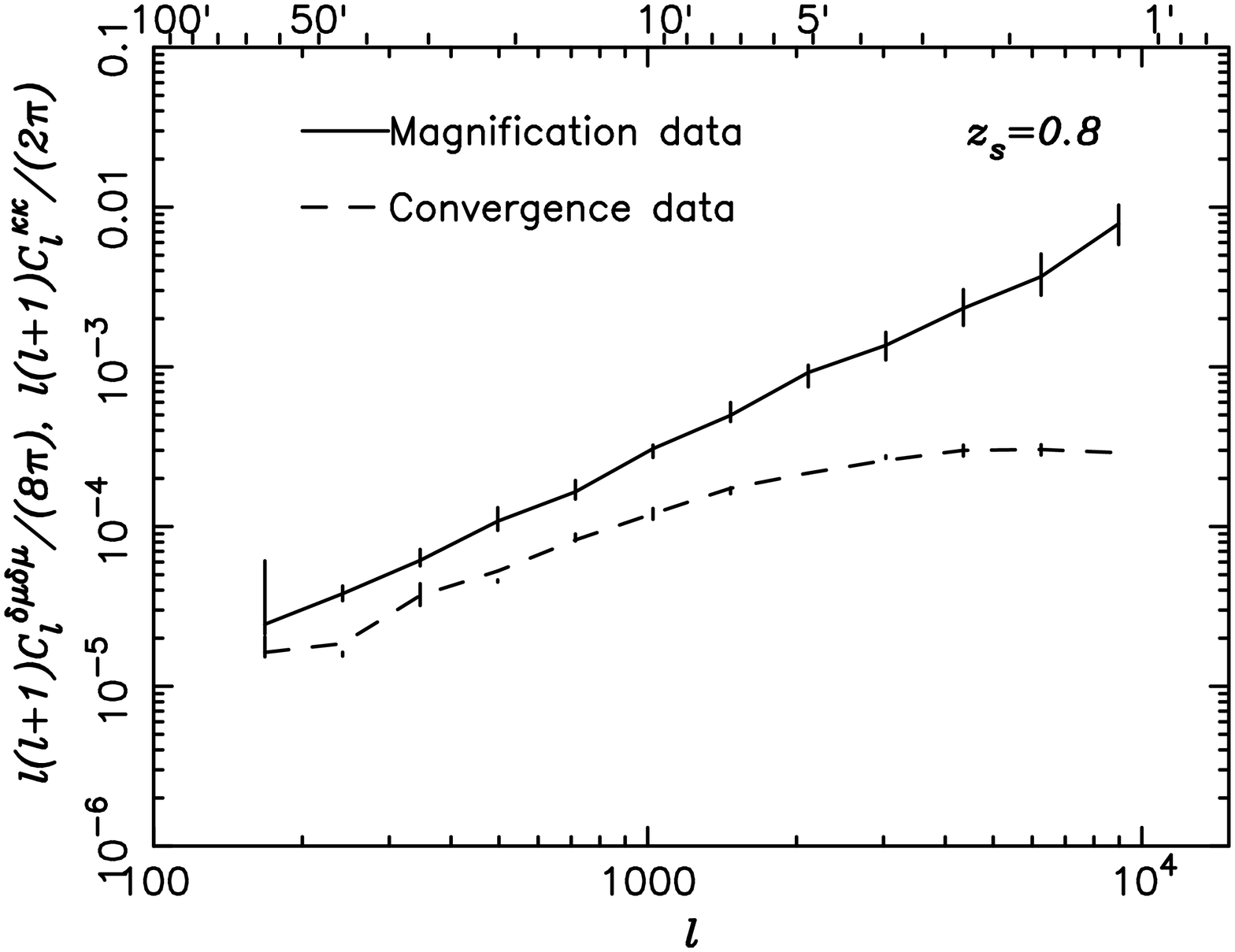,width=8.7truecm,angle=270} }$$ \caption{
Median $\ell(\ell+1)C^{\delta \! \mu\delta \!\mu}_\ell/(8 \pi)$ and
mean $\ell(\ell+1)C^{\kappa\kappa}_\ell/(2 \pi)$ for $z_s = 0.8$. }
\label{pm08}
\end{figure}

\begin{figure}
$$\vbox{
\psfig{figure=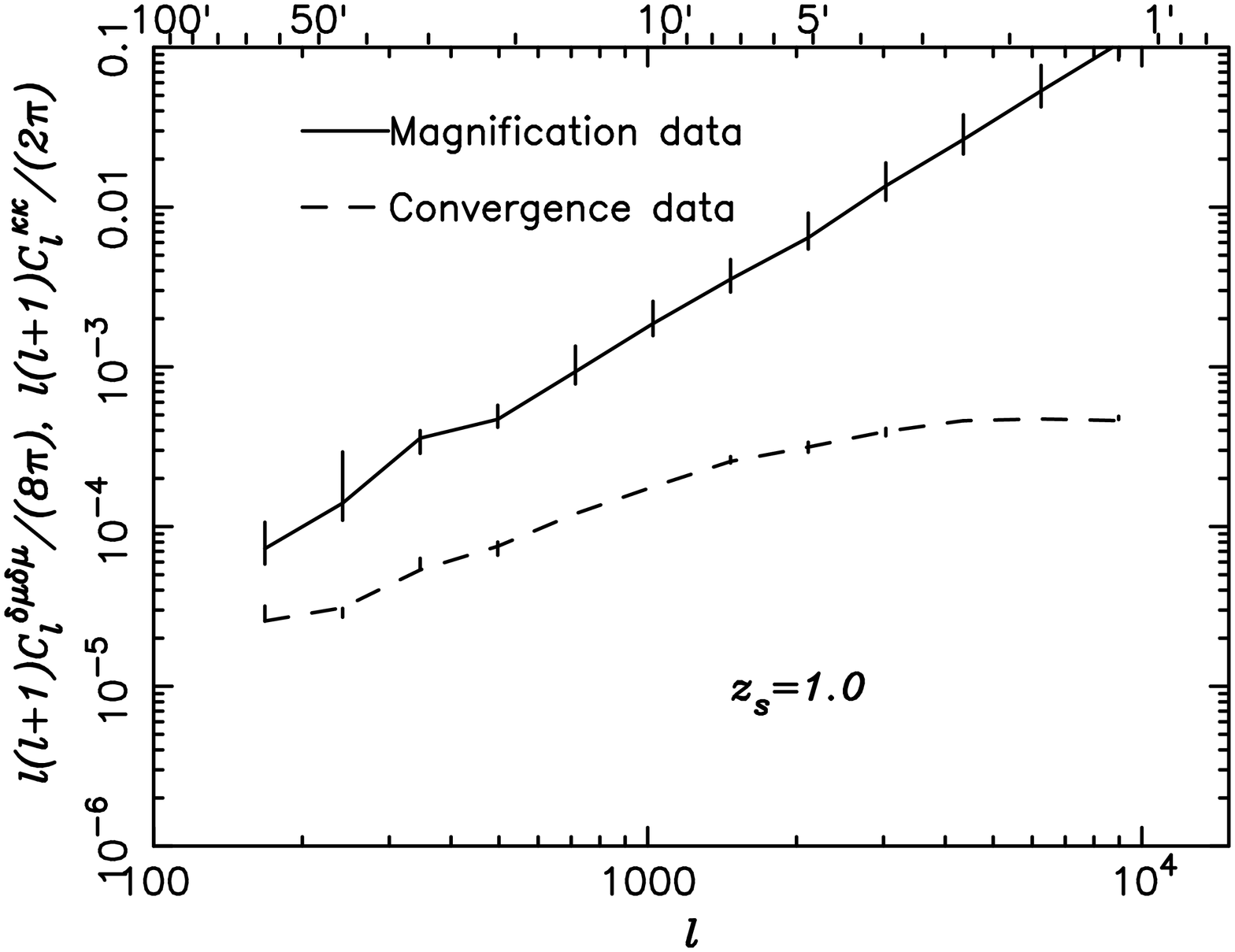,width=8.7truecm,angle=270} }$$ \caption{
Median $\ell(\ell+1)C^{\delta \! \mu\delta \!\mu}_\ell/(8 \pi)$ and
mean $\ell(\ell+1)C^{\kappa\kappa}_\ell/(2 \pi)$ for $z_s = 1$. }
\label{pm1}
\end{figure}

\begin{figure}
$$\vbox{
\psfig{figure=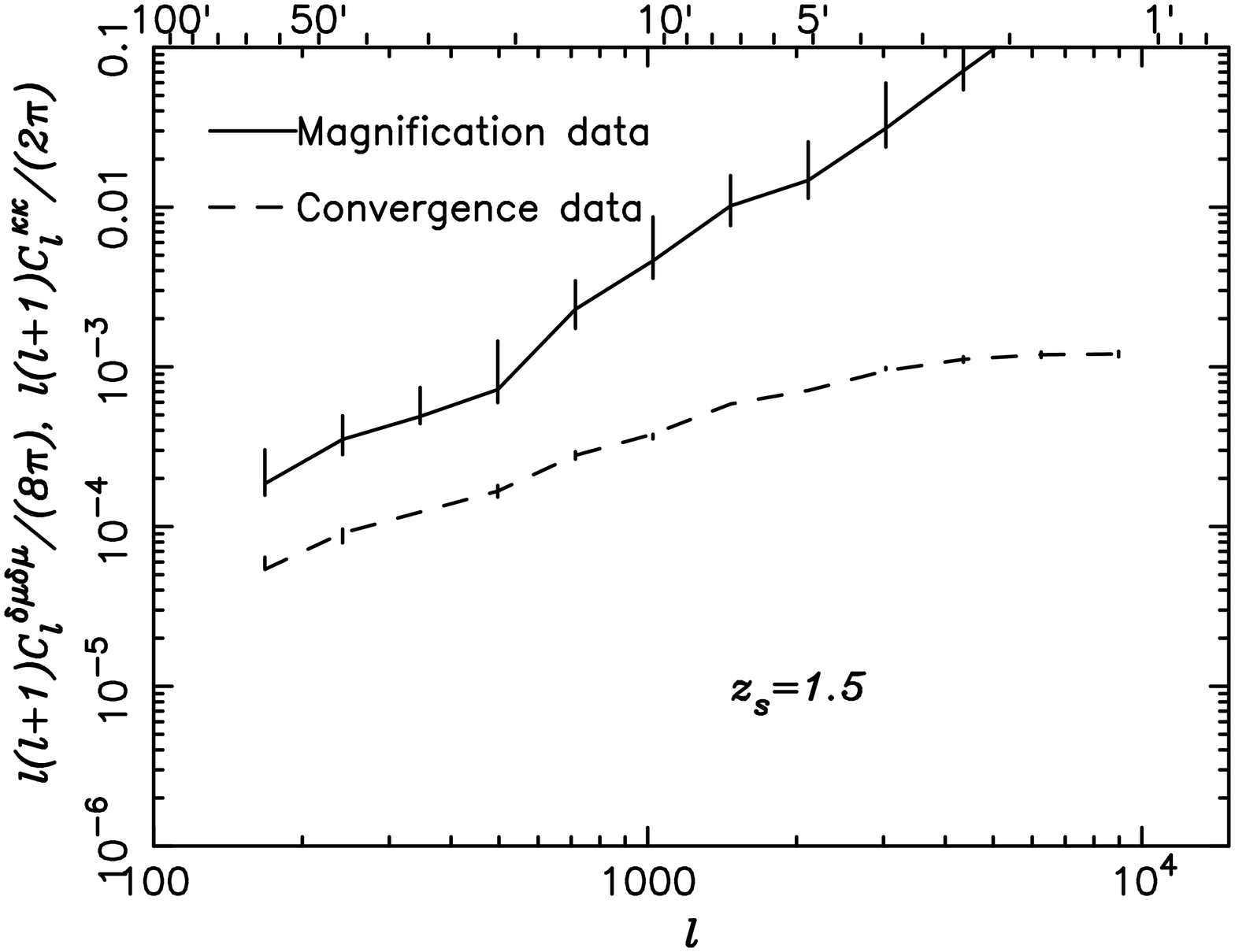,width=8.7truecm,angle=270} }$$ \caption{
Median $\ell(\ell+1)C^{\delta \! \mu\delta \! \mu}_\ell/(8 \pi)$ and
mean $\ell(\ell+1)C^{\kappa\kappa}_\ell/(2 \pi)$ for $z_s = 1.5$. }
\label{pm15}
\end{figure}

\begin{figure}
$$\vbox{
\psfig{figure=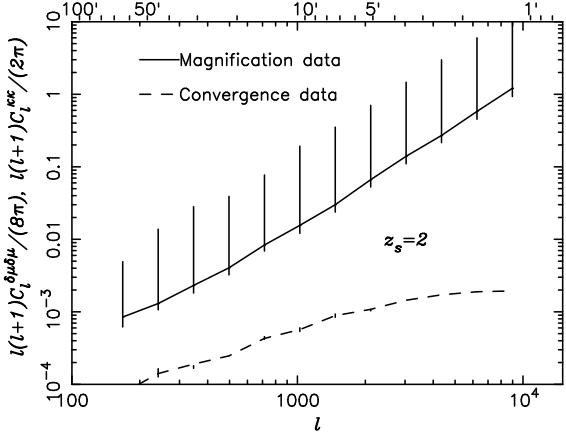,width=8.7truecm,angle=270} }$$ \caption{
Median $\ell(\ell+1)C^{\delta \! \mu \delta \!\mu}_\ell/(8 \pi)$ and
mean $\ell(\ell+1)C^{\kappa\kappa}_\ell/(2 \pi)$ for $z_s = 2$. }
\label{pm2}
\end{figure}

We saw in Section 3.1 that the power in the magnification fluctuation,
$C^{\delta\! \mu \delta\! \mu}_\ell$, is expected to be four times the
shear and the convergence power. This result follows from making the
assumption of weak lensing, where $\delta \mu$ $(\equiv \mu -1)$ is
taken to be equal to $2\kappa$. At low redshift, where the lensing may
be considered to be weak on most scales, $C^{\delta\! \mu\delta
\!\mu}_\ell/4$ and $C^{\kappa\kappa}_\ell$ are consistent with each
other and with the weak lensing approximation. However, as the source
redshift increases, we see increasing departures from this equality.
Figures~\ref{pm05},~\ref{pm08},~\ref{pm1},~\ref{pm15}~and~\ref{pm2}
compare $\ell(\ell+1)C^{\delta\! \mu\delta \! \mu}_\ell/(8 \pi)$ and
$\ell (\ell+1)C^{\kappa\kappa}_\ell/(2 \pi)$ for source redshifts of
0.5, 0.8, 1, 1.5 and 2 respectively. The departures are seen at first
just on small scales (where the lensing is expected to be stronger),
until, finally, at the highest redshifts, the departure is not only
significant, but becomes consistent with a Poisson distribution for
the magnification fluctuations on all scales. In support of this
effect, we noticed the presence of large variations in the
magnification power from one simulation run to the next for sources at
high redshift. Because of these large variations, the curves have been
plotted for the median power values in the magnification, rather than
the mean, which would have been strongly influenced by the large power
values.




\begin{figure}
$$\vbox{
\psfig{figure=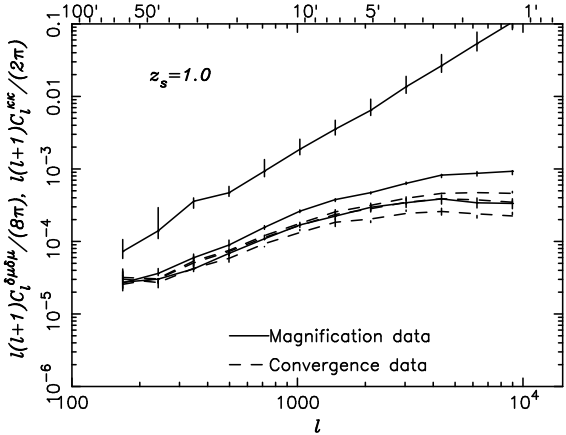,width=8.7truecm,angle=270} }$$
\caption{ 
The three solid curves represent the median $\ell(\ell+1)C^{\delta \!
\mu\delta \!  \mu}_\ell/(8 \pi)$ for the full magnification
distribution (upper curve), and the cut distributions beyond $6\sigma$
(middle solid-line curve) and $1\sigma$ (lower solid-line curve) for
$z_s = 1$. These are compared with the equivalent curves for the mean
$\ell(\ell+1)C^{\kappa\kappa}_\ell/(2 \pi)$ for the convergence
(dotted-line curves).}
\label{powercomb1and6}
\end{figure}

The departures in $C^{\delta \!\mu\delta \! \mu}_\ell/4$ from
$C^{\kappa\kappa}_\ell$ arise from medium and strong lensing
events. By making cuts in the data at the extreme ends, we are able to
show the effects of high magnification events.
Figure~\ref{powercomb1and6} reproduces the magnification and
convergence data from Figure~\ref{pm1} and also shows
$\ell(\ell+1)C^{\delta\!  \mu\delta \!\mu}_\ell/(8 \pi)$ and
$\ell(\ell+1)C^{\kappa\kappa}_\ell/(2 \pi)$ after making cuts in the
data beyond $6\sigma$ and $1\sigma$ in their distributions. When the
extreme magnification (and convergence) events are excluded, we see
that the apparent Poisson distribution in the magnification is removed
and the curve reverts to the more ``normal'' behaviour expected from
lower redshift sources. Thus we can conclude that it is the very high
magnifications which dominate the signal, although there is clear
evidence of departures even for moderate levels of magnification. Our
simulations have a relatively high value for the normalisation,
$\sigma_8$, when compared with very recent determinations (e.g., Brown
et al., 2002b) and we might expect this to lead to a higher number of
clusters than in cosmologies with a lower $\sigma_8$. Consequently,
the shot-noise effect we describe may be moderated, although still
present, in a universe with a lower normalisation value.




\subsection{Higher-order moments}

\begin{figure}
$$\vbox{
\psfig{figure=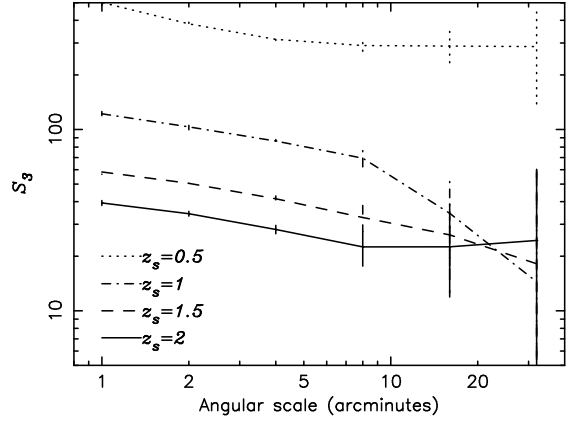,width=8.7truecm,angle=270}
}$$
\caption{
$S_3(\theta,z_s)$ vs. $\theta$ for
$z_s = 0.5$ (dotted line), 1 (dashed-dotted line), 1.5 (dashed line) and 2 (solid
line).
}
\label{s1}
\end{figure}

In Figure~\ref{s1} we plot the $S_3(\theta,z_s)$ statistic for the
convergence, which is a known discriminator of cosmology and
independent of the matter power spectrum normalisation, for four
different source redshifts and for angular scales from $1'.0$ to
$32'.0$.




\begin{figure}
$$\vbox{
\psfig{figure=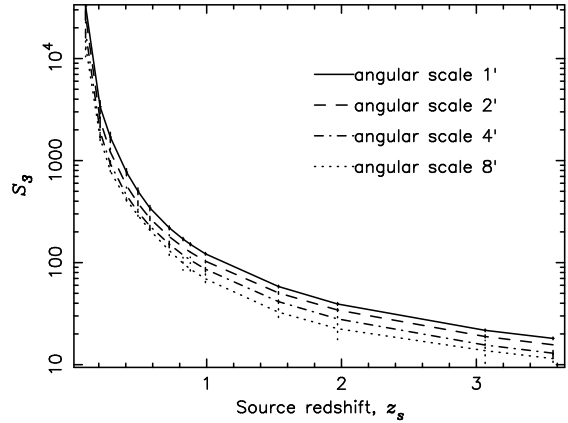,width=8.7truecm,angle=270}
}$$
\caption{
$S_3(\theta,z_s)$ vs. $z_s$ at angular scales of $1'.0$, $2'.0$, $4'.0$ and $8'0$.
}
\label{s2}
\end{figure}

The redshift variation of $S_3(\theta,z_s)$ for the different angular
scales is displayed in Figure~\ref{s2}. On scales larger than
$8'.0$ the errors in the measurements are greater, and these have
not been plotted for clarity. By writing
\begin{equation}
S_3(\theta, z_s) \equiv a(\theta)z^{b(\theta)},
\end{equation}
with $\theta$ expressed in arcminutes, we find
\begin{equation}
a(\theta) = (116 \pm 6) - (7.4 \pm 1.9)\theta
\end{equation}
and
\begin{equation}
b(\theta) = -(2.00 \pm 0.08),
\end{equation}
i.e., $b$ is independent of $\theta$ at a constant value,
so that we may write, without the error values for clarity,
\begin{equation}
S_3(\theta, z_s) \simeq (116 - 7.4\theta)z^{-2.00}
\label{S3approx}
\end{equation}
for $1' \leq \theta \leq 8'$. The fitting to this formula is excellent for all
source redshifts up to and including 1, appropriate to most recent galaxy
surveys. The precise $z_s^{-2.00}$ dependence for $S_3$ is also
particularly simple to apply
when combining or making comparisons with different surveys in which
the median redshifts for the galaxies may be different, making this
relationship particularly valuable.




In Figure~\ref{s3} we have determined the values of
$(S_3\sigma^2_\kappa)(\theta,z_s)$, where $\sigma^2_{\kappa}$
represents the variance in the convergence, for the different source
redshifts. For a given angular scale, this product was found to be
almost independent of redshift for an Einstein-de Sitter universe
by
Bernardeau et al. (1997) and we find a similar
behaviour in the LCDM cosmology here, particularly at high
redshift. Because of the independence from redshift, there is no need
to adjust this statistic when making comparisons amongst different
surveys, provided the median redshifts are not too small. This fact makes
this statistic particularly useful for the discrimination of
cosmologies.

Our results for $S_3 \sigma^2_\kappa$ at low redshift are supported by combining
the shear variance, $\sigma^2_\gamma \equiv
\langle \gamma^2(\theta, z_s)\rangle \approx \sigma^2_\kappa$,
with the expression for
$S_3(\theta, z_s)$ given by equation (\ref{S3approx}).
Barber (2002) has shown that 
\begin{equation}
\langle \gamma^2(\theta, z_s) \rangle = \langle \kappa^2(\theta, z_s)
\rangle \propto z_s^{2.07 \pm 0.04},
\end{equation}
for source redshifts $z_s \leq 1.6$ and angular scales of $2'.0 \leq
\theta \leq 32'.0$, 
so that the combination
\begin{equation}
S_3\sigma^2_\kappa(\theta, z_s) \propto z_s^{2.07 \pm
0.04}.z_s^{-(2.00 \pm 0.08)}
\end{equation}
predicts the very slowly rising function of redshift for $z_s < 1$ and
$2'.0 \leq \theta \leq 8'.0$, which are the ranges of applicability
common to both results.




\begin{figure}
$$\vbox{
\psfig{figure=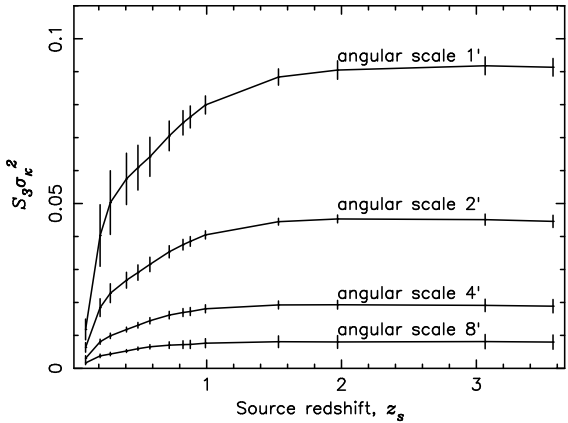,width=8.7truecm,angle=270}
}$$
\caption{
$S_3\sigma_{\kappa}^2$ vs. $z_s$ for angular scales of $1'.0$, $2'.0$, $4'.0$ and $8'.0$.
}
\label{s3}
\end{figure}

\section{DISCUSSION AND CONCLUSIONS}

\subsection{Discussion}

We have shown that the convergence power spectrum values computed
directly from the weak lensing simulations show remarkably good
agreement with the non-linear predictions for the convergence power
based on the 
Smith et al. (2002) Halo Model-inspired fitting formul\ae. Whilst we
have not compared the predictions for other cosmological models in
this paper, the same numerical data, together with equivalent data for
an open cosmology, have been compared with the predictions from
Hierarchical models for the density field. Valageas, Barber \& Munshi
(2003) have shown excellent agreement for the full probability
distribution functions for the shear and Barber, Munshi \& Valageas
(2003) have shown agreement with predictions for the convergence and
higher-order moments. Consequently, these several agreements give us
considerable confidence in our lensing procedure and the resulting
weak lensing data, which we have here analysed in a number of ways.

Our results for the power in the magnification fluctuation suggest
that the observed magnifications should be treated with some care
as an estimate of the convergence power. This is especially true
on small scales and at high redshift, since the magnification is
very sensitive to locally high values of the convergence
(arising from the tail of the skewed convergence distribution) and
may vary significantly across small angular intervals. It further
confirms that the weak lensing regime for the magnification should
be treated with care, even for galaxy surveys with a mean redshift
of 0.8. Worse, at a redshift of 1, where many surveys have been
undertaken or planned, the magnification power may be extremely
noisy. Consequently, evaluation of the convergence from
magnification measurements is clearly likely to be biased by
isolated high peaks in the convergence and shear fields.

The origin of this effect is the appearance of a few regions of
very high magnification in the simulations. The spectrum is
dominated by the strongest magnification events, although the
signal shows moderate departures from the expected values even for
medium magnifications. Since in an LCDM universe clusters form
earlier and have time to relax, we expect large clusters to
generate a large shear field. It is these largest structures with
the highest shears, at the tails of the shear distribution, which
dominate the magnification. These large shear-producing structures
are rare, and so can be assumed to have a Poisson distribution.
Hence the effect is to produce a shot-noise effect in the observed
magnification power spectrum. Since the magnification is a
non-linear function of the shear and convergence, this will not
appear in the individual shear and convergence power spectra, but
should be more apparent in their higher-order moments. Another
consequence of this is that the cosmic variance, from realisation
to realisation, should fluctuate quite wildly, which is the case
here, and motivated the use of median statistics, rather than
means which are more susceptible to this effect.

Since the magnification effect results from lensing from
the largest structures and since the formation of structure evolves at
different rates in different cosmologies, we would expect similar
effects to be present in other cosmologies, but with differing degrees
of severity. We have not studied the effect in other
cosmologies but we might reasonably assume that the LCDM cosmology
would display large effects because of the early development of
structure.

We have shown that the effect can be damped by ``clipping" the
magnification field by removing the highest peaks. Removing the
magnification values beyond 6-$\sigma$ in the distribution removes
most of the effect, showing that indeed it is the highest peaks
causing this effect. One may wonder if this is a purely numerical
effect, as in the simulations there is a high sampling of lines of
sight, including the high-magnification regions around clusters. In
reality we may not expect to see such extreme effects as often since
galaxy surveys only Poisson sample the magnification field. In
addition, if the normalisation, $\sigma_8$, were lower than our chosen
value, the number of clusters would be expected to be smaller, again
leading to fewer examples of extreme magnifications.  However, even
after clipping the magnification values beyond 6-$\sigma$ there is
still a residual systematic. Only after clipping the values beyond
1-$\sigma$ in the distribution do we remove this effect significantly.

This all suggests that estimating the convergence power spectrum
from magnification should be handled with care, unless evaluated
at low redshifts. In addition, the estimation of higher-order
statistics from the magnification will be equally affected by the
shot-noise from individual, massive clusters which will also
increase the effects of sampling variance on any large-scale
measurement of the magnification field. Hence, while magnification
remains a useful tool for probing the mass distributions in
individual clusters, its value as a statistical probe over large
scales may be complicated by these non-linear effects.

The results of
Barber et al. (2000) for $S_3$ on large angular
scales for different cosmological models, and the consistent results
of
Jain
et al. (2000) clearly show the strong $\Omega_m$ dependence of $S_3$.
Consequently,
provided the convergence field is obtained by reconstruction
from the shear, the $S_3$ statistic may be used to good effect to
discriminate different
cosmologies. In determining $S_3$ from their
simulations,
Jain et al. (2000) quote values for sources at redshift 1 only. It is
reassuring to note that the magnitudes of their $S_3$ values on the
different angular scales, determined in quite a different way from
ours (by reconstruction of
the convergence from the shear data), are in good agreement at the
smallest scales where non-linear effects are expected to be most
pronounced. As the angular scale increases our values fall slightly
more rapidly with scale.

Our expression for the angular scale and redshift dependence for $S_3$
clearly demonstrates the necessity of knowing the redshift
distribution of the sources. This is the same conclusion reached by 
Barber
(2002) who reported the angular scale and redshift dependences for the
shear variance.

However, now that the redshift dependence of $S_3$ is established for
our LCDM cosmolgy ($S_3 \propto z_s^{-2.00}$), values determined from
different surveys can be adjusted easily to a fixed median source
redshift, enabling the combination and comparison of data from the
different surveys. The only proviso here is that the redshift
dependence of $S_3$ is likely to be cosmology dependent, since the
magnitude of this statistic on a given angular scale and for a
specific distribution of sources is known to be a discriminator of
cosmologies. We have not yet investigated the redshift dependence of
$S_3$ in other scenarios, although Barber et al. (2003) have reported
on various aspects of $S_3$ from simulations and semi-analytical
predictions in two different cosmologies. We would also hope to
investigate the higher-order statistics for the convergence smoothed
with a compensated filter in a later work, as this will allow more
direct comparison with observed shear data.

The statistic $S_3\sigma^2_\kappa$ for a given angular scale is much
less sensitive to the source redshift, as
Bernardeau et al. (1997) also found for the Einstein-de Sitter
cosmology. Figure~\ref{s3} shows it to be approximately independent of
redshift for high redshift sources and, using the results for the
redshift dependence of the shear variance from
Barber (2002) together with the redshift dependence of $S_3$ reported
here, the statistic is shown to be only a very slowly increasing
function of $z_s$ for $z_s < 1$. Consequently, $S_3\sigma^2_\kappa$
may prove to be very useful for surveys in which the
redshift distribution of the sources is uncertain.

\subsection{Conclusions}

For the convergence power spectrum,
$\ell(\ell+1)C^{\kappa\kappa}_\ell/(2\pi)$, computed numerically from
the lensing statistics in our simulations, we find excellent agreement
with the values computed using a halo model. We can therefore have
considerable confidence in our weak lensing statistics when applied to
the shear, convergence, magnification and higher-order statistics. In
addition, with this consistency, theoretical determinations of lensing
statistics on a wider range of scales than can be achieved with
numerical simulations may be used with increasing confidence.

Our results for one quarter the power in the magnification
fluctuations, $\ell(\ell+1)C^{\delta \! \mu \delta \!\mu}_\ell/(8
\pi)$, show that the magnification is susceptible to the effects
of discrete massive clusters and large variations across small
angular intervals. These effects occur specifically beyond the
weak lensing regime, and become apparent in the magnification even at
low redshift and on small angular scales. Consequently, determination of
the convergence field from magnification data should be treated with special attention.

Our simple mathematical description for $S_3(\theta, z_s)$, showing it
to be closely proportional to $z_s^{-2}$ for the LCDM cosmolgy, makes
it particularly simple to compare and combine the results from
different surveys in which the median redshifts of the galaxies may be
different.

Finally, we found directly, and through combination with the
mathematical expressions for the shear variance and $S_3$, that the
combined statistic $S_3\sigma^2_\kappa$ is only a very slowly
increasing function of redshift for low redshift sources and
approximately independent of redshift at high redshift, as has been
found in the Einstein-de Sitter cosmology. This statistic, therefore,
may be useful in comparing and combining the data from different
surveys with different galaxy redshift distributions.

\section*{ACKNOWLEDGEMENTS}

This work has been supported by PPARC and carried out with facilities
provided by the University of Sussex and the University of
Edinburgh. AJB was supported in part by the Leverhulme Trust. ANT is a
PPARC Advanced Fellow, and thanks the University of Sussex for its
hospitality, where this work began. The original code for the
three-dimensional shear computations was written by Hugh Couchman of
McMaster University.  There were many useful discussions with Ludo Van
Waerbeke, Henk Hoekstra, Peter Schneider, Antonio da Silva, Rachel
Webster, Andrew Liddle, Andrew Melatos, Chris Fluke, Martin White,
Chris Vale and David Bacon.

\end{document}